# A Highly Efficient Hybrid Fiber Optic Laser Using a Cesium Atom Vapor Cell as an Optical Gain Medium

*Seokjin Kim[1], Mingyu Lee[1], Sanggwon Song[1], Seongjin Hong[2], Johan Nilsson[3]\*, and Kyunghwan Oh[1]\**

[1]Photonic Device Physics Laboratory, Department of Physics, Yonsei University, 50 Yonsei-ro, Seodaemun-gu, Seoul 03722, Korea

[2]Depart. of Physics, Chung-Ang University, 47 Heukseok-ro, Dongjak-gu, Seoul, 06974, Korea

[3]Optoelectronics Research Centre, University of Southampton, SO17 1BJ, UK.

E-mail: jn@orc.soton.ac.uk, koh@kyonsei.ac.kr

Keywords: Alkali atom vapor, Atom laser, Laser cavity, Fiber optics

A new scheme of a highly efficient hybrid laser cavity is proposed and experimentally demonstrated utilizing a hot cesium (Cs) vapor cell as an optical gain medium. The laser cavity consists of a macroscopic concave reflecting mirror (>99% reflectivity) and a 4% Fresnel-reflecting perpendicularly cleaved facet of a single mode fiber (SMF). The cylindrical cesium gain cell is located between these two reflectors. The SMF serves multiple roles: 1) a passive mode-matching component to approximate the pump beam diameter to that of the laser cavity mode within the cesium cell, 2) an output coupler with low reflectivity, and 3) a high beam-quality laser delivery with a low loss. Optimizing the pump beam waist diameter and the cesium vapor cell temperature, a high slope efficiency of 86% and continuous wave power of 419 mW were obtained in the pump power range of 400 to 600 mW, with an optical-to-optical conversion efficiency of 71%. The unique multi-functional role of the SMF in the hybrid cavity is fully described, and it can also be applied to other phases of high optical gain media.

## 1. Introduction

Atomic gas lasers featured prominently in early laser physics research [1], but their efficiency was limited by inefficient pumping, such as electrical gas discharge or flash and arc lamps. The recent development of laser diodes (LDs) and their stable high-power operation in array configurations [2, 3] have provided an attractive pumping scheme for a wide range of lasers. LD-pumping has been implemented in various lasers with a variety of optical gain media, including rare-earth-doped fiber lasers [4, 5], solid-state lasers [6, 7], and gas lasers [8]. In particular, rare-earth-doped fiber lasers have been a primary beneficiary of LD pumping for high power and efficiency. This LD-fiber optic combination has led to the development of various high-power fiber lasers that are now used in many industrial, military, and scientific applications [9, 10].

Meanwhile, fiber-coupled optical gain media have been studied in solid, liquid, and vapor phases to take further advantage of the optical fibers' superb light guidance characteristics [11-17]. Optically coupling the macroscopic-scale (~cm in diameter) gain medium to an optical fiber (~10μm in the core diameter) makes it possible to achieve excellent light confinement and guidance control, improving laser efficiency [12, 14] and output power [15, 17]. Optical fibers are also commonly used for low-loss delivery of the laser output, maintaining a well-defined laser spot size, brightness, and polarization over various distances [18]. Continuous development of LDs, optical fibers, and their innovative combinations are still revolutionizing laser technologies [19-22].

Researchers have been further exploring the use of LDs to pump the atomic vapor, which requires pumps with narrow spectral linewidth and wavelength stability [23]. If those requirements are met, alkali atom vapors can show their intrinsic potential to offer a high quantum efficiency [24, 25]. Since the first demonstration of the alkali atom vapor laser [26], various types of laser cavities and pumping schemes have been explored to demonstrate how atomic optical gain can be transformed into practical laser systems. Longitudinal pumping schemes, which use LD arrays in multiple paths and directions, have been studied as a way to increase the absorbed pumping power [27-29]. Stable and unstable cavities with transverse LD pumping schemes have also been investigated for laser power scaling [30-32]. Despite the notable achievement of kilowatt-level output power, the slope efficiency of previous attempts based on LD pumping has been significantly lower than the quantum limit of alkali atoms. These results contrast the slope efficiency achieved with a bulky Ti: sapphire laser pumping in early research [33]. Probable reasons for this discrepancy could be the spatial mismatch between the laser cavity modes and the LD pump beam within the gain medium. This mismatch can limit the conversion of the pump photons, reducing the net optical gain and, subsequently, the laser slope efficiency [25]. Nonlinear ionization processes that result in additional loss around the pump beam could be another attribute of high-power pumping [24].

Recent advances in maximizing the optical gain of alkali-atom vapor have included the adaptation of single mode fibers (SMFs) [34-36]. This approach has successfully matched the spatial distribution of the signal and pump laser beams to maintain a high spatial overlap in the optical gain medium in a single-pass amplifier. The critical merit of using SMF was to guide both the pump and signal in the *LP₀₁* mode since their wavelengths are separated by only ~40 nm to allow the fundamental mode guidance to both beams. This light guidance over SMF resulted in a substantial increase in the amplification factor [35] and output power of the amplifier [36]. However, this spatial beam control via optical fiber has yet to be attempted in lasers, where the laser modes defined by the cavity boundary conditions should be considered. In this study, the authors have introduced a novel fiber optic approach to experimentally demonstrate a highly efficient cesium (Cs) atom vapor laser. They accomplished this by implementing a hybrid cavity (HC), a unique combination of a macroscopic reflector (MR), a microscopic SMF, and a macroscopic atom vapor cell placed between them for the first time. The unique SMF-MR-HC provided new rationales in laser technology: 1)An efficient passive spatial mode-matching: SMF-MR-HC secured the high spatial mode overlap over 90% between the pump and laser cavity mode inside the cesium vapor cell. The proposed scheme successfully approximated the pump beam diameter to that of the laser cavity mode with a stable accuracy within the vapor gain medium, which has not been attempted in previous atom vapor laser cavities. 2) An optimal output coupling: SMF provides a low reflectivity of ~4% Fresnel reflection at the cleaved end facet to satisfy the optimal output coupling ratio for a cesium vapor gain medium, which was not achieved in prior dielectric mirrors. 3) A high beam-quality laser output delivery: The SMF only transmits the laser output in the fundamental *LP01* mode to ensure high beam quality, inherently solving the laser output beam-quality issues in prior technologies.

By optimizing the physical parameters of the hybrid cavity, the authors achieved a significant enhancement in the laser slope efficiency, overcoming one of the oldest hurdles in atom vapor lasers. The report covers the design principles of the hybrid laser cavity, the laser's performance, and a comparison with previous laser cavities.

## 2. Experimental setup

Figure 1 shows a schematic diagram of the experimental configuration. We used an external cavity diode laser (ECDL, Toptica DLC pro) as a pump laser. It had a spectral linewidth of 100 kHz and a tuning range of 12 GHz around $\lambda_P$=852 nm, corresponding to the D₂ transition, $^6S_{1/2}$ → $^6P_{3/2}$ of the cesium atom. See the bottom right inset of Fig. 1. We used an absorption spectroscopy feedback scheme [37], where laser power (~1 µW) tapped from a beam splitter (BS) was monitored as it propagated through a cesium reference cell, "Cs cell-1". The pump laser frequency was locked to the minimum transmission at the F=4 transition. See the inset bottom left. The dichroic mirror (DM) transmits the pump laser at 852 nm and reflects the laser at $\lambda_L$=894 nm, corresponding to the transition, $^6P_{1/2} \rightarrow {}^6S_{1/2}$. "Aspheric lens 1" with a focal length of 13 mm focused the pump laser to single mode fiber (SMF-780) end facet. The SMF was Corning 780HP, whose $LP_{11}$ mode cutoff wavelength was ~730±30 nm, ensuring the fundamental $LP_{01}$ mode guidance for both $\lambda_P$ =852 nm and $\lambda_L$=894 nm, separated only by 40nm in the spectral domain. The fiber end was terminated with an angle to form an anti-reflection (AR) coated angled physical contact (APC) connector, preventing unwanted optical feedback. The other end of the SMF was terminated with an uncoated perpendicular facet and mounted in a standard physical contact (PC) connector. The power of the laser beam was measured by a power meter (Thorlabs, S145C).

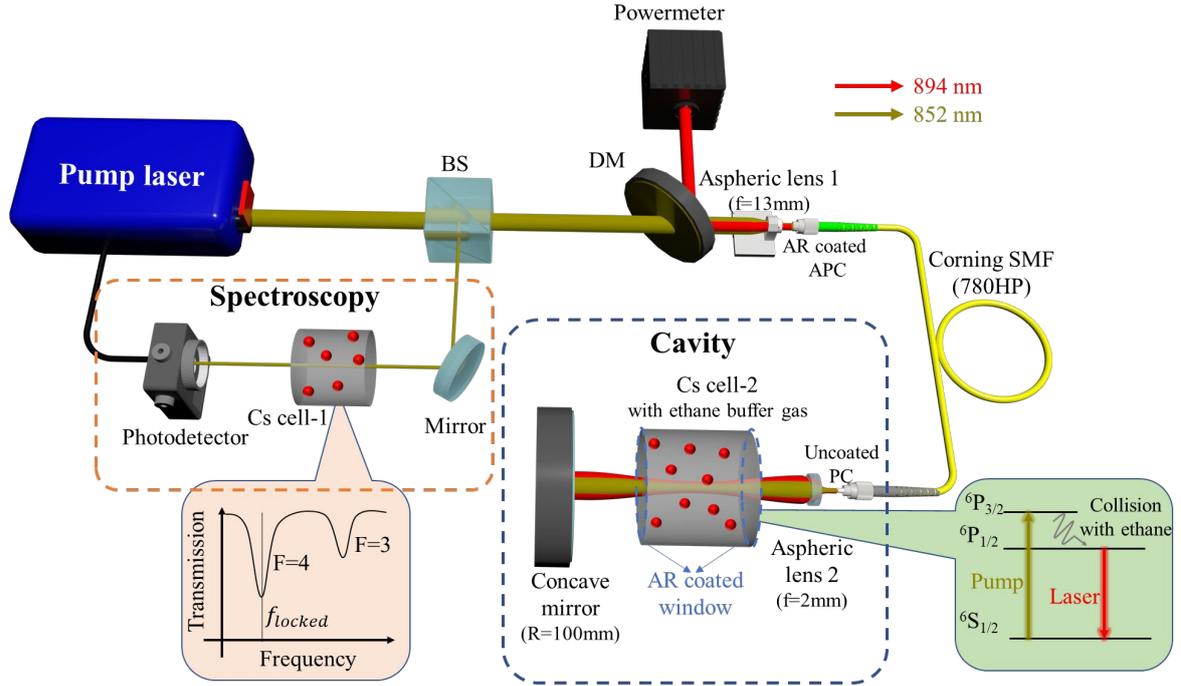

**Fig. 1** The schematic diagram of the experimental setup for hybrid cavity laser. The setup comprises a laser "cavity" and a "spectroscopy". The cavity consists of a concave mirror, and the end facet of single mode fiber (SMF) terminated at an uncoated physical contact (PC) connector. The spectroscopy unit locks the laser frequency to the atomic transition. (BS: beam splitter, DM: dichroic mirror, SMF: single-mode fiber, APC: Angled physical contact connector, PC: Physical contact connector).

Our proposed hybrid cavity consisted of a macroscopic concave spherical mirror with a 100 mm radius of curvature (R~99.7%) and the perpendicularly cleaved facet of SMF-780, whose reflectivity is ~4% due to Fresnel reflection between the silica glass of the optical fiber and the air interface. It uniquely combines the stable high reflectivity of the macroscopic mirror and multi-functional fiber optic output coupler. The optical boundaries, the concave mirror surface and the fiber facet defined the laser modes of our proposed hybrid cavity. The optical gain was provided by "Cs cell-2" between the two cavity mirrors. The frequency-locked pump laser propagated through the SMF-780 in the fundamental $LP_{01}$ mode and exited toward "Cs cell-2", which could be approximated as a Gaussian beam [38, 39]. The pump beam waist diameter was fine-tuned using "Aspheric lens 2" with a focal length of 2 mm with an adjustable position. Note that the surfaces of the aspheric lenses 1 and 2 were all AR-coated to suppress optical feedback. The SMF-780 guided both the laser output at λ=894 nm and the pump at λ=852 nm in its fundamental $LP_{01}$ mode.

The gain medium, "Cs cell-2", formed a cylinder with an axial length of 2 cm and diameter of 2.5 cm and was filled with cesium66 kPa of ethane buffer gas. The buffer gas collision broadens the absorption spectrum of cesium [40] and accelerates the decay from $^6P_{3/2}$ to $^6P_{1/2}$ [41]. The cell windows were AR-coated at both 852 nm and 894 nm. An electric heater enclosing the cell in the lateral direction controlled the vapor temperature from 100 to 140 °C. The laser output was guided in the $LP_{01}$ mode of the SMF and passed through the AR-coated APC. The power meter measured the laser output as it was reflected by the DM.

Optimizing the pump beam waist within the optical gain medium was critical to enhance the laser efficiency since it determines the overall pump absorption and the gain dynamics in the 3-level optical gain system [42] as in the cesium vapor. The pumping rate is insufficient for a collimated pump beam with an excessively large waist. In contrast, when the pump beam is too tightly focused, the pump extends the waist rapidly after the focal point, diminishing the overall pumping rate before it reaches the other end of the cell. By changing the distance between the SMF-780 facet and the aspheric lens 2 with an adjustable fiber collimator (Thorlabs, CFC2-B), the pump beam waist diameter and its axial location were controlled within the optical gain medium, "Cs cell 2".

## 3. Results and discussion

The laser characteristics were measured for various pump beam waists (defined by $1/e^2$ intensity level), and the results are summarized in Fig. 2. Here, we set the "Cs cell 2" temperature at 120 °C. The temperature variation was suppressed under 1°C using a commercial temperature controller. The incident pump power was measured after the "Aspheric lens 2". See Fig. 1. The pump beam waist was located at the center of "Cs cell 2". A beam profiler (Thorlabs, BP104-IR) measured the pump beam waist diameter. The concave mirror location and, subsequently, the cavity length were also optimized for maximum output power. Experimental cavity parameters of the proposed hybrid laser are summarized in Table 1. We kept the reflectance

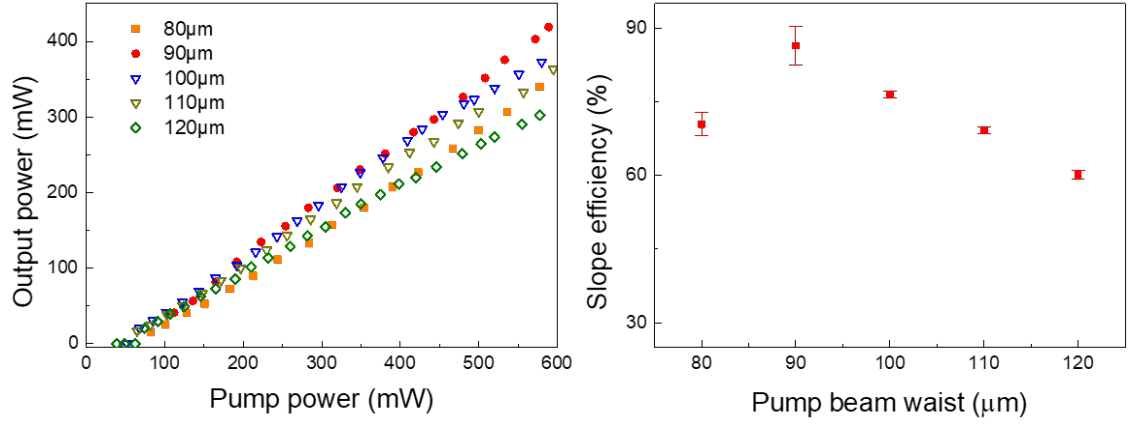

**Fig. 2** a) The output laser power versus the pump power for various pump beam waist diameters. b) The slope efficiency as a function of the pump beam waist diameter. Here the pump beam waist was located at the middle of the "Cs cell 2", whose temperature was maintained 120 °C.

(R~99.7%) and the radius of curvature (100mm) of the concave mirror the same in the experiments. The laser output power reached its maximum of 419 mW at the pump beam waist of 90 μm with 7.5 mm Rayleigh length, and the corresponding pump threshold was 103 mW. Note that the laser power was limited by the available pump laser power.

As the pump beam waist diameter increased further to 120 μm, the laser power reduced to 302 mW. We attributed this behavior to the variation of the pump beam area and its volume overlapping the cesium atoms. Optimally focused pump beam diameter could provide the highest pumping rate, which is inversely proportional to the local pump beam area.[43] See Figure 2 (a). The slope efficiency was plotted as a function of the pump beam waist in Figure 2 (b). We obtained a maximum slope efficiency of 0.865±0.04 at the pump beam waist of 90 μm, and the threshold pump power was 103±28 mW. The slope efficiency decreased to 60.1% as the pump beam waist diameter increased to 120 μm.

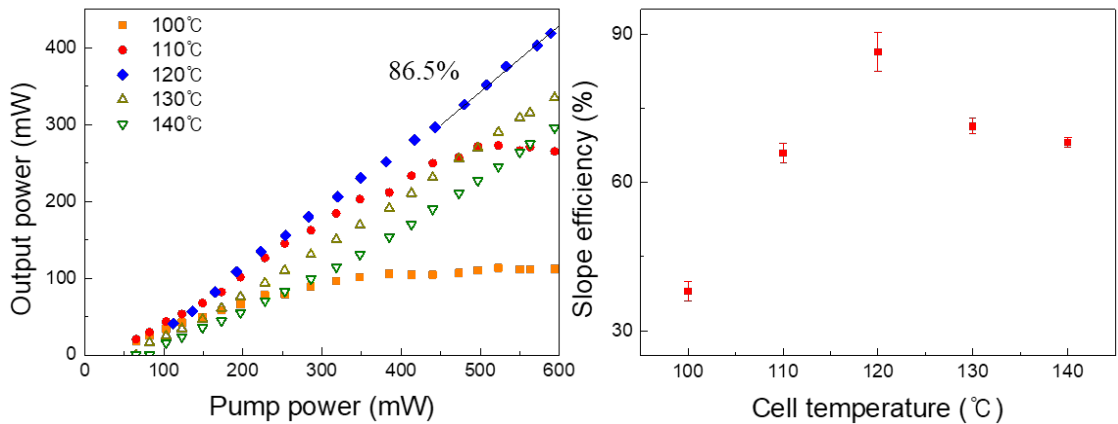

**Fig. 3.** a) The output laser power versus the pump power for various Cs cell temperatures. b) The slope efficiency as a function of the Cs cell temperature. Here the pump beam waist was 90 μm.

| Pump beam waist diameter (μm) | 80 | 90 | 100 | 110 | 120 |
|---|---|---|---|---|---|
| Cavity length (mm) | 133 | 136 | 140 | 143 | 145 |

**Table 1.** Cavity parameters for various pump beam waists.

Note that the pump beam waist diameter was varied using the adjustable collimator, which also varied the cavity length as in Table 1 and, consequently, the laser cavity modes. Therefore, the laser output dependence on the pump beam waist diameter differs from the previous report [42], where the cavity mode was fixed. A non-monotonic dependence of the slope efficiency on the pump beam waist in our experiments indicated that an optimal operation was within the range of the variation in our experiments.

The cesium cell temperature determines the atomic density, partial pressure, and optical gain dynamics [42, 44]. We experimentally optimized the cesium cell temperature and measured its impact on laser output power, summarized in Fig. 3. In Figure 3 (a), we observed a saturation of laser power at the cell temperature of 100 and 110°C. For the cell temperature over 120 °C, the saturation was not observed, but the laser power and threshold significantly varied. They are plotted as a function of the cell temperature in Figure 3 (b). As the temperature rises, the atomic density of cesium also increases, which leads to a higher pump power requirement for achieving population inversion and lasing threshold. The threshold power was measured to monotonically increase from ~15 mW at 100 °C to 162 mW at 140 °C. In contrast, the slope efficiency reached its maximum at the cell temperature of 120 °C and then decreased at higher temperatures.

For the optimal conditions with the pump waist of 90 μm and the cell temperature of 120°C, the proposed hybrid laser provided a slope efficiency of ~86.5% and an optical-to-optical conversion efficiency of 71%, a substantial improvement over prior reports. Table 2 compares our results with some of the previous results.

The alkali atom vapor's high optical gain leads to a low reflectivity of the output coupler [43,45, 46] for optimal output power. Yet, the reflectivity in previous reports was still higher than 30%. Prior laser cavities adopted LD arrays with beam-forming optics to pump alkali atoms in a mixture of multiple high-order Gaussian beams. This inevitably incurred a spatial mismatch between the pump and the laser cavity modes. Our hybrid cavity provided high slope efficiency by 1) securing a stably low reflection in the output coupler using the 4% Fresnel reflection at the interface between the glass fiber facet and air and 2) increasing the spatial overlap between the pump and laser cavity mode.

| Gain medium | Output coupler R(%) | Slope efficiency(%) | O-to-O conversion efficiency(%) | Pump power(W) | Output power(W) | Ref. |
|---|---|---|---|---|---|---|
| K | 60 | 31 | 12 | 41.6 | 5 | 21 |
| Cs | 20 | 52 | 48 | 96 | 48 | 15 |
| Rb | 11 | 53 | 46 | 37 | 17 | 18 |
| Cs | 10 | 61.5 | 49 | 2.9 | 1.4 | 25 |
| Cs | 4 | 86.5 | 71 | 0.589 | 0.419 | This work |

**Table 2.** Comparison of laser characteristics of alkali atom lasers.

In Fig. 4, we schematically showed how the spatial overlap was achieved in our proposed laser. Table 3 calculated the overlap between the pump beam and the fundamental cavity mode along the axial positions using ABCD matrix treatment [47]. We also considered the chromatic aberration of the aspheric lens due to the spectral separation between the pump ($\lambda_P$ =852 nm) and laser ($\lambda_L$=894 nm). The difference in the focal lengths between them was ~2 μm, which was negligible in our experiments. It is noted that the overlap was maintained over 0.92 within the cesium cell, which was not achievable in the prior cavities. The slope efficiency of a laser is mainly determined by four factors as described in the equation below [33]:

$$\frac{dP_L}{dP_P} = \frac{\ln(1-T)}{\ln\{(1-T)(1-L)\}} \frac{\lambda_P}{\lambda_L} \eta_P \frac{dS}{dF} \quad (1)$$

Here we have: $\frac{dP_L}{dP_P}$: slope efficiency, $P_L$ laser power, $P_P$: pump power, $T$: transmission of the output coupler, $L$: total loss of cavity except for the transmission of the output coupler, $\eta_P$:

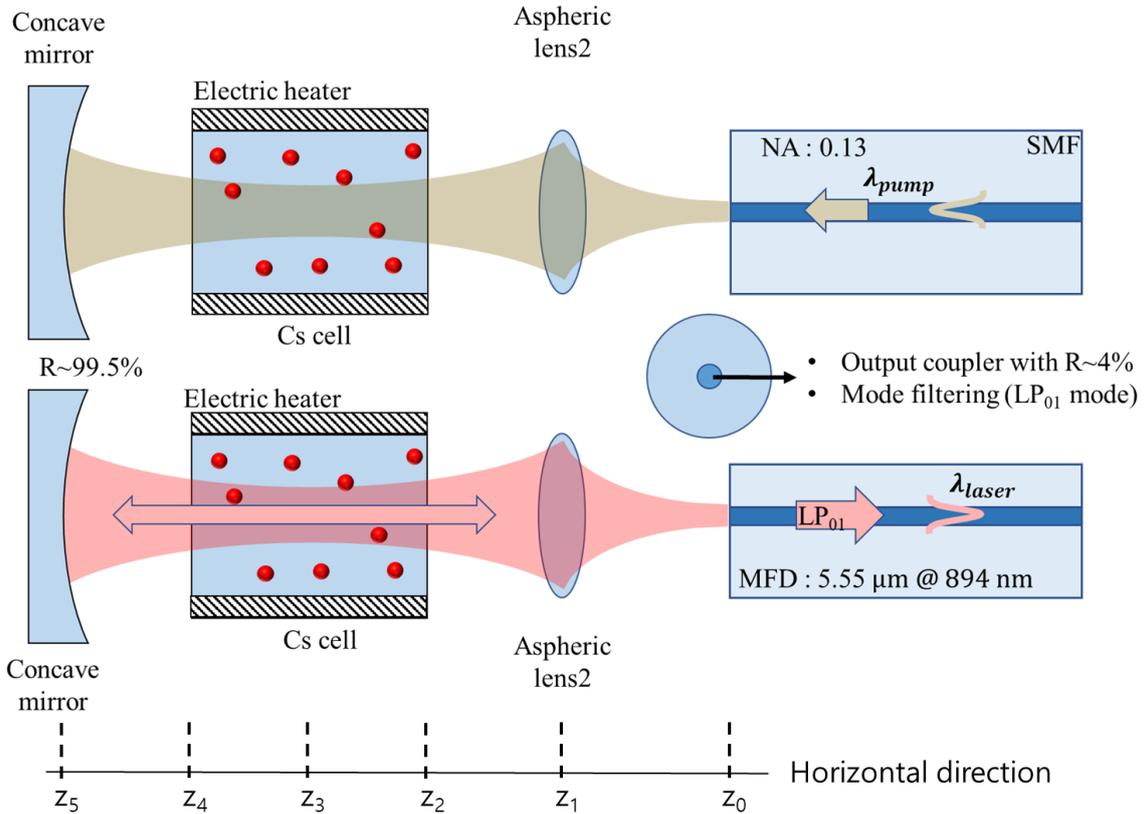

**Fig. 4.** Schematic diagram of the spatial distribution of the pump, and the laser mode in the proposed hybrid DPAL. The upper diagram illustrates the spatial distribution of the pump laser. The lower diagram shows the fundamental cavity mode configuration. At the bottom, the horizontal locations along the cesium cell axis are marked ($z_0$: fiber end facet, $z_1$: Aspheric lens, $z_2$: the right window of Cs cell, $z_3$: the center of Cs cell, $z_4$: the left window of Cs cell, $z_5$: Concave mirror)

|  | $z_5$ | $z_4$ | $z_3$ | $z_2$ | $z_1$ | $z_0$ |
|---|---|---|---|---|---|---|
| $w_P$ (μm) | 601 | 75.2 | 45 | 75.2 | 215.3 | 2.67 |
| $w_L$ (μm) | 607 | 77.3 | 46.8 | 76.2 | 216.9 | 2.78 |
| Overlap | 0.98 | 0.95 | 0.92 | 0.97 | 0.96 | 0.92 |

**Table 3.** Pump laser beam radius ($w_P$), fundamental laser cavity mode radius ($w_L$), and spatial overlap ($w_P^2/w_L^2$) at several axial positions.

ratio of the absorbed pump power over incident pump power, $\frac{dS}{dF}$: efficiency of converted laser photons from absorbed pump photons, $\lambda_P$: pump wavelength, $\lambda_L$: laser wavelength. $T$ and $L$ are in a range from 0 to 1. For the proposed hybrid cavity, $\frac{\lambda_P}{\lambda_L}$ =95.3% $T$=96%, and we estimated $L$=6.7% (loss of aspheric lens=0.2 %, loss of concave mirror=0.3 %, cesium cell window loss=6.2 %) and notably $\eta_P \sim 1$.

Compared to the previous report showing the highest efficiency alkali vapor laser obtained by a bulky Ti:Sapphire laser pumping,[33] $\frac{\ln(1-T)}{\ln\{(1-T)(1-L)\}}$ increased from 0.95 to 0.98, and $\frac{dS}{dF}$ increased from 0.90 to 0.93 in our experiments. Here, $\frac{dS}{dF}$ in our proposed hybrid cavity is the efficiency of the conversion of laser photons in the $LP_{01}$ mode from the absorbed pump photons. By matching the spatial overlap between the laser cavity mode and the pump, especially within the optical gain medium for $z_2 < z < z_4$, as shown in Table 3, to increase $\frac{dS}{dF}$.

The laser output was delivered in the $LP_{01}$ mode of SMF-780, which served as an efficient spatial mode filter for the laser. We placed a beam profiler (Thorlabs, BP104-IR) at the position of the power meter in Figure 1 and measured the beam quality of the laser output operating at the optimal condition. As expected, the $M^2$ of the laser was highly symmetric and ranged from 1.04 to 1.08, comparable to those of low-power fiber lasers [48].

Our study experimentally demonstrated the meaningful advantages of using single mode optical fiber to form a hybrid cavity by increasing the spatial overlap between the pump and laser beams within the gain medium. This method could find applications in laser systems where the pump and laser wavelengths are close in the spectral domain such that both could be guided in the $LP_{01}$ mode of the optical fiber. Therefore, the proposed scheme can be applied to Raman and Brillouin laser with a macroscopic optical gain medium. Power scaling in the proposed hybrid cavity might cause issues, such as optical damage at the optical fiber facet [49] and nonlinear optic effects in the fiber [50-53]. These hurdles might be alleviated using a large-mode-area fiber (LMAF) implemented in high-power laser delivery applications [54, 55], which the authors are pursuing.

## 3. Conclusion

We proposed a new hybrid cavity laser using cesium atom vapor as an optical gain medium. The cavity consisted of a macroscopic reflector with a high reflectivity of 99.7% and a fiber optic output coupler with ~4% Fresnel reflection. The single mode fiber also functioned as a passive mode-matching component to ensure a high overlap (>0.92) between the pump and laser cavity mode was maintained within the cesium optical gain cell. It also facilitated the low coupling losses for the laser to be guided in the LP$_{01}$ mode of SMF-780. Optimizing the hybrid cavity's physical parameters, such as the pump beam width of 90 μm and the cell temperature of 120 °C, we achieved a slope efficiency of 86.5% and an optical-to-optical conversion efficiency of 71%, which is significantly improved in comparison to prior reports. The laser output power was 419 mW at the pump power of 589 mW

**List of Abbreviations**

| | |
|---|---|
| SMF | Single mode fiber |
| LDs | Laser diodes |
| ECDL | External cavity diode laser |
| BS | Beam splitter |
| DM | Dichroic mirror |
| AR | Anti-reflection |
| APC | Angled physical contact |
| PC | Physical contact |
| LMAF | Large-mode-area fiber |

**Availability of data and materials**

The datasets used and/or analysed during the current study are available from the corresponding author on reasonable request.

**Competing interests**

The authors declare that they have no competing interests

**Funding**

This work was supported by the National Research Foundation of Korea (NRF) grant funded by the Korea government (MSIT) (No.2019R1A2C2011293) and institute for information & communications Technology Planning&Evaluation(IITP) grant funded by the Korea government(MIST) (No.2021-0-015610002001,Nanoscale Quantum Emitters Integrated by All-fiber Optofluidics).

**Acknowledgements**

Not applicable.

**Authors' contributions**

SK contributed on the design of the work, acquisition, analysis, and interpret the data, and was a major contributor in writing the manuscript.

ML contributed on the design of the work.